\documentclass[12pt,pra,a4paper,superscriptaddress,onecolumn]{revtex4}
\usepackage[english]{babel}
\usepackage[T1]{fontenc}
\usepackage[utf8]{inputenc}
\usepackage{graphicx,epstopdf}
\usepackage{amssymb}
\usepackage{amsmath}
\usepackage{float}
\usepackage{bbm}
\usepackage{tikz}

\usepackage{amsfonts}
\usepackage{bbm}
\usepackage{color}
\usepackage{latexsym}
\usepackage{caption}
\usepackage{subcaption}
\usepackage{times,txfonts}
\usepackage{color,soul}
\usepackage{listings}
\usepackage{bbold}
\usepackage{mathrsfs}
\usepackage{pgfplots}
\pgfplotsset{compat=1.10}
\usepackage[pdftex,plainpages=false,colorlinks=true,citecolor=blue,linkcolor=blue,urlcolor=blue,filecolor=green,bookmarksopen=true]{hyperref}
\usepackage{hyperref}

\begin{document}
	
	\title{Gravitationally induced wave-function collapse from dynamical bifurcation}
	
	\author{C. A. S. Almeida}
	\email{carlos@fisica.ufc.br}
	\affiliation{Universidade Federal do Cear\'a, Departamento de F\'{i}sica, 60455-760, Fortaleza, CE, Brazil}

	
	\begin{abstract}
		We propose an effective non-relativistic framework in which wave-function collapse emerges as a deterministic dynamical instability induced by gravitational self-interaction and regulated by short-distance repulsion. The dynamics is described by a nonlinear Schr\"odinger equation supplemented by a phenomenological repulsive sector ensuring regularity at high densities. Using a variational Gaussian ansatz, we derive an explicit effective energy functional and show that extended quantum states lose stability beyond a critical mass scale. This loss of stability is associated with a bifurcation in the reduced dynamical system governing the wave-function width, leading to the emergence of stable localized configurations. Within this picture, collapse corresponds to the dynamical selection of one of these localized attractors, driven by infinitesimal asymmetries in the initial state and occurring without stochastic noise or environmental coupling. The mechanism provides a controlled and quantitative realization of gravity-induced localization, extending Schr\"odinger--Newton-type models while avoiding their pathological short-distance behavior. Possible implications for mesoscopic systems probing the quantum-to-classical transition are briefly discussed.

			\end{abstract}

	\maketitle
	
\section{Introduction}

The emergence of classical behavior from quantum mechanics remains a central problem at the interface between quantum theory and gravitation. While the linear and unitary evolution governed by the Schr\"odinger equation has been extensively validated at microscopic scales, the apparent absence of macroscopic quantum superpositions raises the question of whether additional physical mechanisms become relevant when gravitational effects associated with mass distributions are taken into account. 
In this work, ``extended quantum states'' refer to spatially delocalized 
wave functions characterized by a large width parameter $\sigma$, in 
contrast to localized states with finite spatial extent.

Within the standard framework, the suppression of macroscopic coherence is usually attributed to environmental decoherence \cite{Zurek2003}. Although decoherence successfully explains the dynamical damping of interference terms in reduced density matrices, it does not by itself induce genuine state reduction nor select a unique outcome. This limitation has motivated the exploration of alternative mechanisms capable of dynamically destabilizing extended quantum superpositions.

Environmental decoherence induced by gravitational or relativistic effects has been extensively discussed in the literature, including models based on time-dilation and gravitationally induced phase dispersion \cite{Blencowe,Pikovski}. While such mechanisms suppress interference, they do not by themselves generate dynamical attractors or genuine state localization.

Gravity-induced collapse models constitute a particularly well-motivated class of proposals in this context. Early arguments due to Di\'osi \cite{Diosi1987} and Penrose \cite{Penrose1996,Penrose1998} suggest that quantum superpositions of distinct mass distributions may possess an intrinsic gravitational instability, characterized by a timescale determined by the corresponding gravitational self-energy. These ideas have inspired both stochastic collapse models and deterministic semiclassical approaches. A comprehensive review of such models, together with their phenomenology and experimental constraints, is provided in \cite{Bassi2013}. More recent developments have significantly refined both the theoretical and experimental status of collapse models, particularly in the context of non-interferometric tests and mesoscopic systems (see, e.g., \cite{Carlesso2022,Bassi2023,Bilardello2019}).

Among deterministic formulations, the Schr\"odinger--Newton equation has received considerable attention as a semiclassical description of gravitational self-interaction sourced by the quantum probability density itself \cite{Moroz1998,Giulini2011}. While this approach captures a natural tendency toward localization, it is purely attractive and generically leads to pathological short-distance behavior, including unbounded collapse and the absence of stable ground-state configurations.

Stochastic gravity-induced collapse models, on the other hand, introduce noise terms designed to enforce localization dynamically. Although such models are internally consistent, they are subject to increasingly stringent experimental bounds, particularly from precision measurements probing macroscopic quantum coherence \cite{Donadi2021}. These constraints motivate renewed interest in deterministic and noise-free alternatives.

Recent work has further explored deterministic gravity-induced collapse mechanisms based on non-local gravitational self-energy contributions \cite{JusufiSingletonLobo}. While such approaches emphasize non-locality at the level of the gravitational interaction, the present work adopts a complementary effective perspective, focusing on the dynamical instability of extended quantum states and the emergence of localization through a bifurcation mechanism.

In this work, we adopt a deliberately minimal and effective approach. We consider a non-relativistic quantum system subject to gravitational self-interaction, supplemented by a phenomenological short-distance repulsive sector. The purpose of the repulsive contribution is not to introduce a new fundamental interaction, but to regularize the dynamics at high densities and ensure the existence of stable localized configurations. The resulting evolution is governed by a nonlinear Schr\"odinger equation that should be interpreted as an effective description, valid below relativistic scales and above microscopic distances where a full theory of quantum gravity would be required.

Our central result is that, within this effective framework, extended quantum states generically lose stability beyond a critical mass scale. Using a variational formulation, we derive an explicit effective energy functional and show that the competition between kinetic dispersion, gravitational self-attraction, and short-distance repulsion leads to a bifurcation in the reduced dynamical system governing the wave-function width. Above the critical threshold, stable localized solutions emerge, while extended configurations become dynamically unstable.

Within this picture, wave-function collapse is identified with the deterministic evolution toward one of these localized attractors. The dynamics is strictly deterministic; however, effective unpredictability arises from sensitivity to arbitrarily small asymmetries in initial conditions,
 driven by the amplification of arbitrarily small asymmetries in the initial state. No stochastic noise, environmental coupling, or ad hoc collapse postulates are invoked. The mechanism extends Schr\"odinger--Newton-type dynamics by providing a controlled short-distance regularization and a quantitative criterion for the onset of localization.
 
In parallel, considerable progress has been made in proposals to probe quantum aspects of gravity using mesoscopic systems, including entanglement-based tests and optomechanical platforms \cite{Bose2017,Marletto2017,Carney2021}. These developments highlight the increasing relevance of gravitational effects in the quantum-to-classical transition regime.

The paper is organized as follows. In Section 2 we present the effective non-relativistic framework and perform a variational analysis of the nonlinear dynamics, deriving a critical instability threshold. Section 3 discusses the role of short-distance regularization and the associated physical scales. In Section 4 we interpret the resulting instability as a dynamical collapse mechanism and contrast it with decoherence and stochastic collapse models. Section 5 contains our conclusions and outlook.

\section{Variational analysis and dynamical instability}

In order to provide a more systematic foundation for the effective dynamics, 
we introduce an effective Lagrangian density
\begin{equation}
\mathcal{L} = \frac{i\hbar}{2}(\psi^* \dot{\psi} - \psi \dot{\psi}^*) 
- \frac{\hbar^2}{2m}|\nabla \psi|^2 
- \frac{1}{2}\rho V_{\text{grav}}[\rho] 
- \frac{\lambda}{2} \rho^2,
\end{equation}
where $\rho = |\psi|^2$ and the gravitational potential is given by
\begin{equation}
V_{\text{grav}}(r) = -Gm \int d^3r' \frac{\rho(r')}{|r-r'|}.
\end{equation}


\subsection{Effective energy functional}

We consider a gravitationally modified nonlinear Schr\"odinger equation
\begin{equation}
i\hbar \frac{\partial \psi}{\partial t}
=
\left[
-\frac{\hbar^2}{2m}\nabla^2
+
V_{\mathrm{grav}}[\rho]
+
V_{\mathrm{rep}}[\rho]
\right]\psi,
\end{equation}
with $\rho(\mathbf{r}) = |\psi(\mathbf{r})|^2$.
The corresponding conserved energy functional reads
\begin{equation}
E[\psi]
=
T[\psi] + E_{\mathrm{grav}}[\rho] + E_{\mathrm{rep}}[\rho],
\end{equation}
where
\begin{align}
T[\psi] &= \frac{\hbar^2}{2m}\int d^3r\, |\nabla\psi|^2,\\
E_{\mathrm{grav}}[\rho] &= -\frac{Gm^2}{2}\int d^3r\,d^3r'\,
\frac{\rho(\mathbf{r})\rho(\mathbf{r}')}{|\mathbf{r}-\mathbf{r}'|}.
\end{align}
For definiteness, we consider a local repulsive contribution of the form
\begin{equation}
E_{\mathrm{rep}}[\rho] = \frac{\lambda}{2}\int d^3r\, \rho^2(\mathbf{r}),
\qquad \lambda>0,
\end{equation}
which penalizes high-density configurations and regularizes the short-distance behavior.

\subsection{Gaussian ansatz}

To proceed analytically, we adopt a normalized Gaussian trial wave function,
\begin{equation}
\psi(\mathbf{r}) =
\frac{1}{(\pi\sigma^2)^{3/4}}
\exp\!\left(-\frac{r^2}{2\sigma^2}\right),
\end{equation}
where $\sigma$ characterizes the effective spatial width of the quantum state.

Substituting this ansatz into the energy functional yields
\begin{align}
T(\sigma) &= \frac{3\hbar^2}{4m\sigma^2},\\
E_{\mathrm{grav}}(\sigma) &= -\frac{Gm^2}{\sqrt{2\pi}\,\sigma},\\
E_{\mathrm{rep}}(\sigma) &= \frac{\lambda}{(2\pi)^{3/2}\sigma^3}.
\end{align}
The effective energy as a function of $\sigma$ is therefore
\begin{equation}
E(\sigma)
=
\frac{3\hbar^2}{4m\sigma^2}
-
\frac{Gm^2}{\sqrt{2\pi}\,\sigma}
+
\frac{\lambda}{(2\pi)^{3/2}\sigma^3}.
\label{E_sigma}
\end{equation}

\subsection{Stability and critical threshold}

Stationary configurations correspond to extrema of $E(\sigma)$, determined by
\begin{equation}
\frac{dE}{d\sigma}=0.
\end{equation}
For sufficiently small mass $m$, the kinetic and repulsive terms dominate, and the energy possesses a single stable minimum at large $\sigma$, corresponding to an extended quantum state.

As $m$ increases, the gravitational contribution becomes increasingly important. Beyond a critical mass $m_c$, the extended minimum loses stability, and new minima at finite $\sigma$ emerge. The instability threshold is determined by the simultaneous conditions
\begin{equation}
\frac{dE}{d\sigma}=0,
\qquad
\frac{d^2E}{d\sigma^2}=0,
\end{equation}
which define a saddle-node bifurcation in the reduced dynamics, corresponding 
to the appearance of additional extrema in the effective energy $E(\sigma)$ 
and the associated change in stability of stationary configurations.

At the critical point, the conditions $dE/d\sigma = 0$ and 
$d^2E/d\sigma^2 = 0$ imply the coalescence of two stationary points of 
$E(\sigma)$, which is the characteristic signature of a saddle-node 
bifurcation in one-dimensional gradient systems.

Solving these equations yields a critical mass scale of the order
\begin{equation}
m_c \sim \left(\frac{\hbar^2}{G\lambda^{1/2}}\right)^{1/3},
\end{equation}
up to numerical factors of order unity. While the precise value of $m_c$ depends on the phenomenological parameter $\lambda$, the existence of a finite instability threshold is robust.

\paragraph*{Dimensionless formulation of the bifurcation.}
It is instructive to rewrite the effective energy in
dimensionless form.  Introducing the rescaled width
$\tilde{\sigma}=\sigma/\sigma_0$, with
$\sigma_0 \equiv \hbar^2/(Gm^3)$, Eq.~(\ref{E_sigma}) becomes
 \begin{equation}
  \tilde{E}(\tilde{\sigma})
 = \frac{A}{\tilde{\sigma}^2} - \frac{B}{\tilde{\sigma}}+ \frac{C}{\tilde{\sigma}^3},
\end{equation}
where $A$, $B$, $C>0$ depend only on $m$, $\lambda$, and
fundamental constants.  Solving $d\tilde{E}/d\tilde{\sigma}=0$
and $d^2\tilde{E}/d\tilde{\sigma}^2=0$ simultaneously gives a
critical dimensionless width $\tilde{\sigma}_c = \mathcal{O}(1)$,
confirming that the saddle-node bifurcation occurs at a natural scale in the rescaled problem.  For $m<m_c$ the energy has a single minimum at large $\tilde\sigma$; at $m=m_c$ a saddle-node
appears; for $m>m_c$ a stable localized minimum and an unstable
extremum emerge, in direct correspondence with Fig.~1.

\subsection{Interpretation as a dynamical bifurcation}


The reduced dynamics for the width parameter $\sigma(t)$ can be obtained 
within a time-dependent variational framework. Promoting $\sigma$ to a 
dynamical collective coordinate and inserting the Gaussian ansatz into the 
effective Lagrangian, one obtains an equation of motion of the form
\begin{equation}
M_{\rm eff} \ddot{\sigma} = -\frac{dE}{d\sigma},
\end{equation}
where $E(\sigma)$ is the effective energy functional derived above and 
$M_{\rm eff}$ is an effective inertial parameter associated with the 
variational dynamics.

Here $M_{\rm eff}$ is an effective variational inertial
parameter arising from the time-dependent Gaussian ansatz
and should not be interpreted as a new physical mass scale;
its value depends on the choice of variational family and
does not affect the qualitative structure of the bifurcation.

In realistic situations, coupling to unresolved degrees of freedom, internal 
modes, or weak environmental interactions can lead to effective dissipation. 
At a phenomenological level, this can be incorporated by introducing a damping term,
\begin{equation}
M_{\rm eff} \ddot{\sigma} + \gamma \dot{\sigma} = -\frac{dE}{d\sigma},
\end{equation}
where $\gamma > 0$ is an effective damping coefficient.

In the overdamped regime, $\gamma \gg M_{\rm eff}$, the inertial term becomes 
negligible, and the dynamics reduces to a first-order equation,
\begin{equation}
\dot{\sigma} = -\Gamma \frac{dE}{d\sigma}, 
\qquad \Gamma = \frac{1}{\gamma}.
\end{equation}

The overdamped regime should be interpreted as a coarse-grained effective limit in which fast microscopic degrees of freedom have been integrated out. In this sense, the first-order equation provides an effective description of relaxation toward stationary configurations, rather than a fundamental dynamical law.

It should be emphasized that the bifurcation described here is established within a reduced variational framework based on a Gaussian ansatz. While this does not constitute a proof of bifurcation at the level of the full functional space of wave functions, it provides a controlled and physically transparent indication of an instability mechanism in the underlying dynamics.

This reduced equation should be understood as an effective coarse-grained 
description capturing relaxation toward extrema of the energy functional. 
It arises as the overdamped limit of the second-order dissipative dynamics 
and is analogous to gradient-flow equations commonly encountered in 
dissipative systems such as Ginzburg--Landau dynamics.

Within this framework, the stability properties of the system are directly 
controlled by the structure of $E(\sigma)$. For masses below the critical 
threshold, the extended configuration corresponds to a stable fixed point of 
the reduced dynamics. At the critical mass, this fixed point loses stability, 
and new stable branches corresponding to localized configurations emerge. 

The transition can thus be interpreted as a bifurcation in the phase portrait 
of the reduced dynamical system. Wave-function collapse is identified with 
the deterministic evolution toward one of the stable localized attractors 
that appear beyond the critical point, driven by the amplification of 
infinitesimal asymmetries in the initial state.

Importantly, the dynamics remains strictly deterministic; however, effective 
unpredictability arises from sensitivity to arbitrarily small asymmetries in 
the initial conditions. No stochastic noise or external randomness is required 
at the level of the effective description.

\section{Short-distance regularization and physical scales}

A purely attractive self-interaction generically drives localization without bound, leading to pathological behavior at arbitrarily short length scales. This feature is well known in deterministic self-gravitating models such as the Schr\"odinger--Newton equation, where the absence of a stabilizing mechanism prevents the existence of well-defined stationary states \cite{Moroz1998,Giulini2011}. Any consistent effective description of gravity-induced localization must therefore include a mechanism that counteracts gravitational self-attraction in the high-density regime.

In the present framework, this role is played by an explicit short-distance repulsive sector. The repulsive contribution is introduced phenomenologically and should not be interpreted as a new fundamental interaction. Rather, it parametrizes effective non-local or coarse-grained gravitational effects that become relevant below a characteristic length scale. As shown in Section~2, the inclusion of this repulsive term renders the effective energy functional bounded from below and ensures the existence of finite-width stationary configurations.

In this sense, the parameter $\lambda$ should be interpreted as an effective parameter encoding short-distance physics beyond the present model. Its role is analogous to phenomenological regularization scales introduced in other collapse models, such as the correlation length in CSL or the smearing scale in the Di\'{o}si-Penrose framework.

In particular, non-local extensions of gravity -- such as super-renormalizable quantum gravity \cite{Modesto2012} and ghost-free infinite-derivative theories \cite{Biswas2012} -- provide well-motivated ultraviolet completions in which short-distance gravitational behavior is naturally regularized by non-local form factors. The repulsive sector introduced here may be viewed as an effective parametrization of analogous non-local effects in the semiclassical regime.

The relation to Schr\"odinger--Newton dynamics is particularly transparent in this formulation. In the limit where the repulsive coupling is taken to zero, the effective energy functional reduces to that of the Schr\"odinger--Newton system, and the pathological collapse behavior is recovered. The present model may therefore be regarded as a minimal regularized extension of Schr\"odinger--Newton dynamics, in which short-distance repulsion stabilizes the localization process and allows for a controlled dynamical analysis.

The repulsive sector introduces an effective length scale that characterizes the onset of regularization. This scale should be understood as an emergent or coarse-graining length, rather than as a new fundamental constant. No direct identification with the Planck length is assumed. Depending on the physical context, it may be associated with the scale at which gravitational self-interaction becomes effectively non-local, with smearing effects arising from quantum spacetime fluctuations, or with characteristic lengths appearing in Di\'osi-Penrose-type analyses.

Related attempts to regularize or extend Schr\"{o}dinger-Newton-type dynamics by incorporating non-local gravitational effects have been recently discussed in \cite{JusufiSingletonLobo}. In contrast to those approaches, the present framework emphasizes the role of a dynamical instability regulated at short distances, leading to a bifurcation structure in the reduced dynamics.

The variational analysis presented in Section~2 provides a quantitative criterion for the onset of instability of extended states. The critical mass scale,
\begin{equation}
m_c \sim \left(\frac{\hbar^2}{G \lambda^{1/2}}\right)^{1/3},
\end{equation}
defines the threshold beyond which gravitational attraction overcomes kinetic dispersion and repulsive regularization. While the precise numerical value of $m_c$ depends on the phenomenological parameter $\lambda$, the existence of a finite critical mass is a robust feature of the model. For reasonable choices of $\lambda$, the instability occurs in a mesoscopic regime, well above microscopic scales and far below astrophysical ones, suggesting potential relevance for experimental platforms probing the quantum-to-classical transition.

Although the parameter $\lambda$ is introduced phenomenologically, one may 
estimate its order of magnitude by requiring that the repulsive term becomes 
relevant at length scales where gravitational self-interaction would otherwise 
lead to unphysical collapse. This suggests that $\lambda$ effectively encodes 
short-distance regularization at mesoscopic scales, and its value can be 
treated as a phenomenological parameter constrained by consistency with 
experimental bounds on macroscopic quantum coherence.

In this sense, the short-distance regularization does not merely cure a mathematical inconsistency but plays a central physical role: it enables the emergence of stable localized configurations while preserving the effective character of the description and avoiding unphysical ultraviolet behavior.

\paragraph*{Order-of-magnitude estimates.} To make the physical content of $\lambda$ and $m_c$ more
concrete, we note that the repulsive term $E_{\rm rep}\sim
\lambda/\sigma^3$ becomes comparable to the gravitational term
$E_{\rm grav}\sim Gm^2/\sigma$ at the regularization length
$\sigma\sim\ell_{\rm reg}$, giving
\begin{equation}
\lambda \;\sim\; Gm^2\,\ell_{\rm reg}^2.
\label{eq:lambda_est}
\end{equation}
Substituting into $m_c\sim(\hbar^2/G\lambda^{1/2})^{1/3}$ yields
\begin{equation}
m_c \;\sim\;
\left(\frac{\hbar^2}{G\,\ell_{\rm reg}}\right)^{1/3}.
\end{equation}
For $\ell_{\rm reg}\sim 10^{-7}$~m (a mesoscopic smearing scale
compatible with the Di\'{o}si--Penrose framework) this gives
$m_c\sim 10^{-17}$~kg, in the nano- to micro-mechanical regime
accessible to current optomechanical and matter-wave
interferometry experiments.  These estimates should be
understood as order-of-magnitude guides; the precise value of
$m_c$ depends on $\ell_{\rm reg}$, which is constrained by
consistency with experimental bounds on macroscopic quantum
coherence.

From this perspective, the present model can be viewed as a minimal regularized extension of Schr\"{o}dinger-Newton dynamics, addressing its well-known short-distance instabilities while preserving its effective semiclassical character.

\section{Dynamical collapse as a bifurcation phenomenon}

The competition between kinetic dispersion, gravitational self-attraction, and short-distance repulsion gives rise to a nontrivial dynamical structure in the space of quantum states. As demonstrated by the variational analysis of Section~2, the stability of a given configuration is controlled by parameters such as the total mass and the effective spatial width of the wave function.

Within the reduced description in terms of collective variables, extended quantum states correspond to fixed points of the effective dynamics governing the wave-function width. For sufficiently weak gravitational self-interaction, these fixed points are stable, and the evolution remains close to that predicted by linear quantum mechanics. In this regime, small perturbations decay in time, and macroscopic superpositions are dynamically allowed.

As the control parameters are varied, a critical threshold is reached at which the extended configuration loses stability. Beyond this point, the reduced dynamical system undergoes a bifurcation: the previously stable extended fixed point becomes unstable, and new stable branches corresponding to localized configurations emerge. This transition is a direct consequence of the structure of the effective energy functional and does not rely on any external stochastic element.

The loss of stability of the extended configuration corresponds to a bifurcation in the reduced dynamical system governing the wave-function width. This behavior is schematically illustrated in Fig.~1, which represents the qualitative structure of the effective dynamics derived from Eq.~(18). The diagram is 
intended to illustrate the qualitative structure of the phase portrait and 
does not represent a numerical solution. Note that the schematic bifurcation diagram for the reduced dynamics of the 
wave-function width $\sigma$, as described by the effective first-order 
equation $\dot{\sigma} = -\Gamma \, dE/d\sigma$, obtained in the overdamped 
limit of the dissipative dynamics discussed in Sec.~2.4. 
\begin{figure}[t]
\centering
\begin{tikzpicture}
\begin{axis}[
    width=0.8\linewidth,
    height=0.45\linewidth,
    axis lines=left,
    xlabel={Control parameter (mass $m$)},
    ylabel={Wave-function width $\sigma$},
    xmin=0, xmax=1.1,
    ymin=0.2, ymax=1.1,
    xtick={0.5},
    xticklabels={$m_c$},
    ytick=\empty,
    samples=100,
    domain=0:1,
    clip=false
]

\addplot[
    thick
]
{0.9};

\addplot[
    thick,
    dashed,
    domain=0.5:1
]
{0.9};

\addplot[
    thick,
    domain=0.5:1
]
{0.5 - 0.3*sqrt(x-0.5)};

\addplot[
    thick,
    domain=0.5:1
]
{0.5 + 0.3*sqrt(x-0.5)};

\addplot[
    thin,
    dashed
]
coordinates {(0.5,0.15) (0.5,1.1)};

\node at (axis cs:0.25,0.95) {stable};
\node at (axis cs:0.75,0.95) {unstable};
\node at (axis cs:0.75,0.45) {stable};

\end{axis}
\end{tikzpicture}
\caption{
Schematic bifurcation diagram for the reduced dynamics of the 
wave-function width $\sigma$. For masses below the critical value $m_c$, the system exhibits a single 
stable fixed point corresponding to an extended quantum state. At $m = m_c$, 
this fixed point loses stability. For $m > m_c$, stable localized branches 
emerge, while the extended configuration becomes dynamically unstable. 
Solid (dashed) lines denote stable (unstable) fixed points. }
\label{fig:bifurcation}
\end{figure}
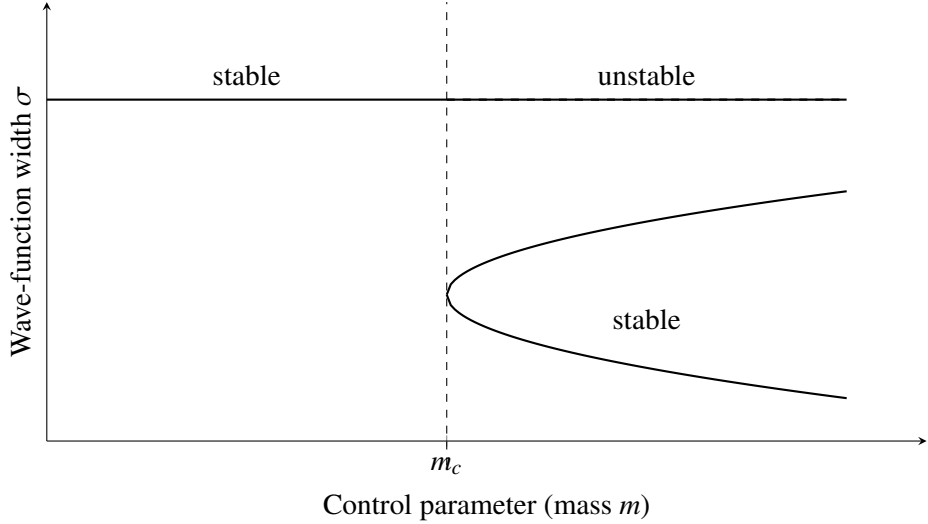

Within this picture, wave-function collapse is identified with the deterministic evolution toward one of the localized attractors. The dynamics remains strictly deterministic. However, effective unpredictability arises from sensitivity to arbitrarily small asymmetries in the initial conditions. The selection of a specific localized configuration is driven by the amplification of arbitrarily small asymmetries in the initial state, which are unavoidable in any physical realization. Importantly, no stochastic noise, environmental coupling, or measurement-induced randomness is introduced at the level of the effective dynamics.

On the other hand, the underlying change in the structure of the effective energy 
$E(\sigma)$ is shown in Fig.~2. This change in the structure of the energy landscape underlies the 
bifurcation in the reduced dynamics described in Sec.~2.4. The curves 
are schematic and intended only to illustrate the qualitative features 
of the effective energy functional.
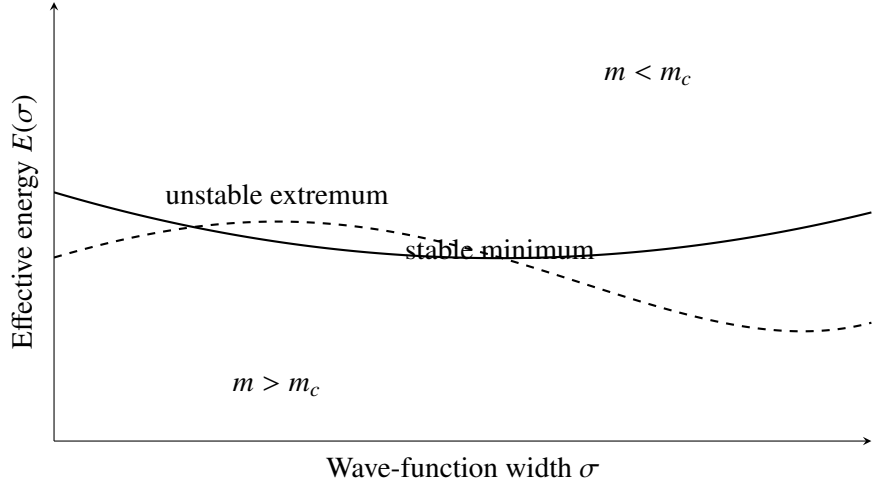
\begin{figure}[t]
\centering
\begin{tikzpicture}
\begin{axis}[
    width=0.75\linewidth,
    height=0.45\linewidth,
    axis lines=left,
    xlabel={Wave-function width $\sigma$},
    ylabel={Effective energy $E(\sigma)$},
    xmin=0, xmax=2.2,
    ymin=-1.2, ymax=1.2,
    xtick=\empty,
    ytick=\empty,
    samples=200,
    domain=0:2.2,
    clip=true
]
\addplot[
    thick
]
{0.25*(x-1.2)^2 - 0.2};

\addplot[
    thick,
    dashed
]
{0.15*(x-0.6)^4 - 0.6*(x-0.6)^2};

\node at (axis cs:1.6,0.8) {$m < m_c$};
\node at (axis cs:0.6,-0.9) {$m > m_c$};

\node at (axis cs:1.2,-0.15) {stable minimum};
\node at (axis cs:0.6,0.15) {unstable extremum};

\end{axis}
\end{tikzpicture}
\caption{
Qualitative behavior of the effective energy $E(\sigma)$ as a 
function of the wave-function width $\sigma$, as defined in Eq.~(12). 
For $m < m_c$, the energy exhibits a single minimum at large $\sigma$, 
corresponding to a stable extended quantum state. At the critical mass 
$m = m_c$, this minimum becomes marginally stable. For $m > m_c$, the 
energy develops additional extrema, including stable localized minima at 
finite $\sigma$ and an unstable extremum separating them, in agreement 
with the conditions $dE/d\sigma = 0$ and $d^2E/d\sigma^2 = 0$ discussed 
in Sec.~2.3.
}
\label{fig:bifurcation}
\end{figure}

It is instructive to contrast this mechanism with environmental decoherence and with stochastic collapse models. Decoherence suppresses interference terms in reduced density matrices through entanglement with external degrees of freedom but does not modify the underlying unitary structure of the dynamics nor generate true attractors in state space. Stochastic collapse models, on the other hand, introduce noise terms designed to enforce localization dynamically, but are subject to increasingly stringent experimental constraints \cite{Donadi2021}. In contrast, the present mechanism achieves localization through a deterministic instability of the effective dynamics itself, without invoking noise or violating existing experimental bounds.

\paragraph*{Collapse timescale.}
A rough estimate of the collapse timescale can be obtained by
linearising $\dot\sigma = -\Gamma\,dE/d\sigma$ around the
unstable extremum $\sigma_*$.  Writing
$\sigma(t)=\sigma_*+\delta\sigma(t)$ one finds
\begin{equation}
\delta\dot\sigma
\;=\; \Gamma\bigl|E''(\sigma_*)\bigr|\,\delta\sigma,
\end{equation}
so the system departs from the unstable configuration
exponentially with characteristic time
\begin{equation}
\tau_{\rm collapse}
\;\sim\;
\frac{1}{\Gamma\,\bigl|E''(\sigma_*)\bigr|}.
\label{eq:tau}
\end{equation}
For $m\sim m_c\sim 10^{-17}$~kg and
$\sigma_*\sim\ell_{\rm reg}\sim 10^{-7}$~m, the curvature
$|E''(\sigma_*)|\sim Gm^2/\sigma_*^3\sim 10^{-10}$~J/m$^2$.
Taking $\Gamma\sim 1/(m\omega)$ with a characteristic
internal frequency $\omega\sim 10^3$~rad/s (typical for
nano-mechanical oscillators) yields
$\tau_{\rm collapse}\sim 10^{-6}$--$10^{-3}$~s, overlapping
with coherence times in current optomechanical platforms.
A full numerical study of $\sigma(t)$ is left for future work.

The bifurcation-based interpretation of collapse provides a natural and physically transparent account of the quantum-to-classical transition in self-gravitating systems. Microscopic systems lie deep within the stable extended regime, where quantum superpositions persist. As the relevant control parameters increase, the system crosses the instability threshold, and localization becomes dynamically inevitable. While the present analysis is formulated within a single-particle effective framework, the mechanism is expected to be further enhanced in many-body systems, where gravitational self-interaction scales with the collective mass density.

In the present framework, the mass parameter $m$ should be understood as an 
effective collective mass associated with the relevant degree of freedom. 
In measurement-like situations, this may include contributions from both the 
system and its immediate environment, reflecting the effective mass density 
entering the self-gravitational interaction. The present model does not aim 
to provide a complete measurement theory, but rather a dynamical instability 
mechanism that may become relevant when such effective mass exceeds a critical 
threshold.

We emphasize that the present mechanism does not constitute a complete solution to the quantum measurement problem. Instead, it provides a dynamical instability that may contribute to the emergence of localization in regimes where gravitational self-interaction becomes relevant.

In summary, the emergence of localized states in the present model is not imposed by postulate but arises as a consequence of the nonlinear structure of the effective self-gravitating dynamics. The resulting collapse mechanism is deterministic, quantitatively controlled, and compatible with the effective character of the description.

\section{Discussion and conclusions}

We have presented an effective framework in which wave-function collapse emerges as a deterministic dynamical bifurcation induced by gravitational self-interaction and regulated by short-distance repulsion. The collapse mechanism does not rely on stochastic noise, environmental decoherence, or external postulates.

By clearly separating the effective dynamics from fundamental quantum field theories, the present approach avoids conflicts with quantum electrodynamics and experimental constraints on noise-driven collapse models. The interpretation of collapse as a bifurcation provides a coherent and physically transparent effective description.

It is important to distinguish the present mechanism from conventional environmental decoherence \cite{Blencowe,Pikovski}. While decoherence suppresses interference terms in the reduced density matrix due to entanglement with external degrees of freedom, it does not by itself select a unique outcome nor induce genuine state reduction. In contrast, the collapse mechanism discussed here arises from an intrinsic dynamical instability of the effective self-gravitating dynamics and persists even in the absence of an external environment. In this sense, gravitationally induced bifurcation may be viewed as complementary to environmental decoherence, providing a dynamical route to state localization that decoherence alone cannot account for.

Although the present analysis is formulated at the level of an effective single-particle wave function, the extension to many-particle systems is conceptually straightforward. In a multi-particle setting, the relevant quantity governing gravitational self-interaction is the collective mass density, which naturally enhances the instability mechanism as the total mass increases. One therefore expects the bifurcation behavior discussed here to become increasingly pronounced for macroscopic systems, providing a natural route toward classical localization in many-body contexts. A detailed treatment of such extensions is left for future work.

Several directions for future investigation naturally emerge from the present work. 
A more detailed numerical investigation of the time-dependent dynamics 
associated with the reduced equation for $\sigma(t)$ would provide further 
insight into the collapse process and timescales. This will be addressed in 
future work. These include the derivation of quantitative estimates for the critical mass and length scales associated with the bifurcation, the exploration of possible experimental signatures in platforms probing macroscopic quantum coherence such as optomechanical systems or matter-wave interferometers \cite{Arndt,Aspelmeyer}, and the analysis of connections with gravitational mean-field formulations in weakly relativistic regimes.  As a matter of fact, recent experimental and phenomenological analyses have placed increasingly stringent constraints on noise-driven collapse models \cite{Donadi2021}, further motivating the exploration of deterministic and noise-free alternatives such as the one discussed here. Addressing these issues would further clarify the physical scope and testability of the proposed framework.

\section{Acknowledgment}

The author express their gratitude to Funda\c{c}\~{a}o Cearense de Apoio ao Desenvolvimento Cient\'{i}fico e Tecnol\'{o}gico (FUNCAP) and Conselho Nacional de Desenvolvimento Cient\'{i}fico e Tecnol\'{o}gico (CNPQ) for their invaluable financial support. He 
acknowledges the support from the grant CNPq/309553/2021-0 (CNPQ/PQ) and by Project UNI-00210-00230.01.00/23 (FUNCAP).

\end{document}